\documentclass[conference,a4paper]{IEEEtran}
\IEEEoverridecommandlockouts
\usepackage{cite}
\usepackage{amsmath,amssymb,amsfonts}
\usepackage{algorithmic}
\usepackage{graphicx}
\usepackage{textcomp}
\usepackage{xcolor}
\usepackage[utf8]{inputenc}

\usepackage{url}
\usepackage{subcaption}
\usepackage{graphicx}
\usepackage{scrextend}
\usepackage[colorlinks=true, allcolors=blue]{hyperref}
\usepackage{tikz}
\usepackage{tkz-tab}
\usepackage{multirow}
\usepackage{listings}
\usepackage{latexsym}
\usepackage{amssymb}
\usepackage{amsmath}
\usepackage{subcaption}
\usepackage{mathtools}
\usepackage[]{moresize}
\usepackage{soul}
\usepackage{enumitem}   
\usepackage{subfiles}

\usepackage{tabularx,booktabs}
\newcolumntype{C}{>{\centering\arraybackslash}X} % centered version of "X" type
\newcolumntype{b}{X}
\newcolumntype{s}{>{\hsize=.5\hsize}X}
\newcolumntype{v}{>{\hsize=.3\hsize}X}

\def\BibTeX{{\rm B\kern-.05em{\sc i\kern-.025em b}\kern-.08em
		T\kern-.1667em\lower.7ex\hbox{E}\kern-.125emX}}
		
\newcommand{\newver}[1]{#1}

\usepackage[
top    = 1.9cm,
bottom = 2.90cm,
left   = 1.4cm,
right  = 1.4cm]{geometry}

\begin{document}
	
%%%%%%%%%%%%%%%%%%%%%%%%%%%%
%% METADATA
%%%%%%%%%%%%%%%%%%%%%%%%%%%%

\title{Cloud-native 5G experimental platform with over-the-air transmissions and end-to-end monitoring} 
% Previous titles:
% - Cloud-native monitoring of MEC 5G networks with Open5GS and Amarisoft \sergio{[Experimental Evaluation]}

\author{\IEEEauthorblockN{Sergio Barrachina-Muñoz, Miquel Payaró, and Josep Mangues-Bafalluy}
	\IEEEauthorblockA{\textit{Services as Networks (SaS)} \\
		\textit{Centre Tecnològic Telecomunicacions Catalunya (CTTC/CERCA)}\\
		Barcelona, Spain \\
		\{sergio.barrachina, miquel.payaro, josep.mangues\}@cttc.cat}
}

\maketitle

% Enable page numbering
% \thispagestyle{plain}
% \pagestyle{plain}

%%%%%%%%%%%%%%%%%%%%%%%%%%%%
%% ABSTRACT
%%%%%%%%%%%%%%%%%%%%%%%%%%%%
\begin{abstract}
	
	% Intro 5G and beyond
	5G represents a revolutionary shift with respect to previous generations given its design centered on network softwarization. Within such a change of paradigm, cloud-native solutions are widely regarded as the future of vertical application development because of their enhanced flexibility and adaptability to complex and dynamic scenarios.
	% This paper
	In this context, we present an experimental framework with over-the-air transmissions that tackles two critical aspects for enhancing the lifecycle management of 5G and beyond networks: cloud-native deployments of 5G core network functions (NFs) and end-to-end monitoring.
	% Contributions
	First, we deploy Open5GS and Prometheus-based monitoring as containerized network functions (CNFs) in a Kubernetes cluster spanning a multi-tier network with a multi-access edge computing (MEC) host. We then demonstrate the end-to-end monitoring system by showcasing via Grafana dashboards both infrastructure resources and radio metrics of two scenarios; one devoted to user plane function (UPF) re-selection and the other to user mobility.
	
\end{abstract}

\begin{IEEEkeywords}
	5G, cloud-native, Open5GS, Kubernetes, monitoring, MEC, experimental platform
\end{IEEEkeywords}

%%%%%%%%%%%%%%%%%%%%%%%%%%%%
%% ToC (TO BE REMOVED)
%%%%%%%%%%%%%%%%%%%%%%%%%%%%
% \sergio{*** ToC is TO BE REMOVED ***}
% \tableofcontents

%%%%%%%%%%%%%%%%%%%%%%%%%%%%
%% INTRODUCTION
%%%%%%%%%%%%%%%%%%%%%%%%%%%%
\section{Introduction} \label{section:introduction}

% 5G in the context of MARSAL
5G and beyond (B5G) networks are set to address the demands of a fully connected and mobile society, enabling a wide variety of services and applications over the same infrastructure. Further, 5G adopts edge computing as a key paradigm evolving from centralized architectures towards multiple points-of-presence (PoPs) of edge nodes. In turn, this enables multi-access edge computing (MEC) applications for low-latency and high bandwidth like virtual and augmented reality (VR/AR).
% MARSAL
In this context, projects like MARSAL~\cite{marsal,vardakas2021towards} propose a new paradigm of elastic virtual infrastructures that integrate transparently a variety of novel radio access, networking, management, and security technologies to deliver end-to-end transfer, processing, and storage services in an efficient and secured way.

% Cloud native 
To materialize the MARSAL vision in 5G and B5G networks, two elements are key: cloud-native deployments within the network function virtualization (NFV) paradigm and end-to-end monitoring. As for the former, cloud-native infrastructures make it possible to share infrastructure resources, enabling their dynamic allocation to meet the service level agreements (SLAs) of existing and future demanding use cases. Cloud-native technologies also reduce time to market, respond sooner to customer demands, and facilitate the lifecycle management and automation of the network. Therefore, they are widely regarded as the future of vertical application development with enhanced flexibility, scalability, and reduced cost.

% Need of monitoring
Monitoring is the second key aspect this paper deals with, as it is critical to manage such complex cloud-native infrastructures and to increase operational efficiency. Indeed, a paramount factor for 5G is end-to-end real-time monitoring, gathering infrastructure metrics (i.e., compute, storage, and network) as well as domain-specific metrics of components such as gNBs or MEC services. Gathering these metrics supports the lifecycle management of services running over the 5G network and favors intelligent reconfiguration and alerting to involved tenants and stakeholders, spanning from infrastructure owners, operators, slice owners, or service/application developers. Certainly, multi-tenant networks must support network slices, which require monitoring key performance indicators (KPIs), commonly belonging to different technological domains and managed by different entities. For instance, a network operator shall focus on the KPIs of the components running the network slice (e.g., RAN, cloud/edge, routers), while the slice owner may consider high-level KPIs (e.g., end-to-end delay) for SLA validation~\cite{mekki2021scalable}. In both cases, monitoring turns into an imperative aspect.

Even though there are valuable works in the literature on cellular networks devoted to cloud-native deployments and monitoring, only a few of them (partially) treat both aspects together. In contrast, this paper deals with both aspects by presenting a cloud-native experimental platform for MEC-enabled 5G networks endowed with an end-to-end monitoring system. The platform is easily deployable via Helm charts.\footnote{All of the source code of our experimental platform~\cite{barrachina2022zenodo} is open.} So, the main contributions are as follows:
\begin{itemize}
    \item Cloud-native 5G core deployment with Open5GS in a multi-PoP Kubernetes cluster, where each NF runs as a separate containerized NF (CNF).
    \item End-to-end containerized monitoring gathering both infrastructure and radio/RAN metrics via CNFs.
    \item Integration of a commercial gNB (Amarisoft Callbox) into the 5G testbed and development of a custom sampling function for pulling gNB metrics (e.g., downlink/uplink bitrate).
    \item Showcasing of the monitoring system through the visualization of different metrics in Grafana dashboards. Two toy scenarios are considered; one for UPF re-selection and the other for UE mobility.
\end{itemize}

% Structure
The rest of the article is structured as follows. Section \S\ref{section:row} discusses the related work, section \S\ref{section:platform} describes the main components of the experimental platform and the MEC-enabled testbed, section \S\ref{sec:monitoring} depicts the end-to-end monitoring system, and section \S\ref{section:evaluation} showcases and validates the whole framework. Finally, we our conclusions and future work is collected in section \S\ref{section:conclusions}.

%a holistic cloud-native monitoring system based on Kubernetes to support 5G system in the form of CNFs. We use Open5GS as an open-source 5G core and integrate it with Amarisoft Callbox, a commercial gNB providing the RAN. We use Prometheus and Grafana to gather and plot the metrics of NFs and host nodes, respectively. Besides, we show in a toy scenario the convenience of MEC regarding delay-criticall use cases.

%%%% SoA - Papers not included

%Authors in \cite{wiranata2020automation} proposes network automation using Kubernetes to manage 5G services. TERRIBLE PAPER. CONS: Mosaic and OAI in Kubernetes but whole core network in one pod. No monitoring

%Authors in \cite{chima2020context} propose a custom Kubernetes scheduler for edge-native 5G applications 5G. CONS: 5G NSA (unknown project) with no cloud-native deployment of 5GC NFs. No monitoring.

%\cite{bolivar2018deployment} depicts an architectural blueprint of 5G-aware open-source evaluation testbed. CONS: no 5G core discussed nor deployed, and not monitoring.

%\cite{perez2021performance} presents a performance comparison of well-known virtualization technologies applied to deploy a real monitoring architecture for multi-site 5G platforms. CONS: No core deployed + RAN monitoring is not considered.

%%%%%%%%%%%%%%%%%%%%%%%%%%%%
%% RELATED WORK
%%%%%%%%%%%%%%%%%%%%%%%%%%%%
\section{Related work} \label{section:row}
There are several works in the literature dealing either with cloud-native 5G deployments or end-to-end monitoring systems for 5G networks. However, none of them fully addresses both aspects together as we discuss next.

% Cloud-native 5Gcore
As for cloud-native deployments, a method to orchestrate and manage a container-based C-RAN using Kubernetes and OpenAirInterface (OAI) is presented in~\cite{novaes2020virtualized}. Authors in~\cite{haavisto2019open} introduce an open-source infrastructure for 5G RAN development where DevOps simplifies the deployment of end-to-end applications to the edge. Also, Kube5G is proposed in~\cite{arouk2020kube5g} for building and packaging a cloud-native telco NF through nested layers, and a 5G cloud-native environment based on Kubernetes and Openshift Operator is introduced in~\cite{arouk20205g}. Recently, an integration of KubeFed for deploying workloads in multiple clusters and Network Service Mesh for providing connectivity across cluster boundaries has been proposed in~\cite{osmani2021multi}. The aforementioned works make valuable efforts towards cloud-native deployments. However, none of them realizes a full 5G core, but different variations of 4G's evolved packet core (EPC). Further, they do not focus on monitoring.

% Monitoring
As for papers dealing with monitoring, authors in \cite{trakadas2018scalable} present SONATA, a multi-PoP monitoring framework of NFV services involving both containers and virtual machines. However, monitoring is discussed from an architectural perspective, and no actual 5G core is deployed. Instead, authors in \cite{giannopoulos2021monitoring,giannopoulos2021holistic} present an approach to deliver monitoring and telemetry mechanisms as a service using Prometheus and Netdata over an Open5GS network, but no containerization is provided. Similarly, a monitoring framework introducing metrics collectors deployed per network slice using Prometheus is devised in~\cite{mekki2021scalable}, but no containerized deployment of the 5G core is provided either. In \cite{goshi2021investigating}, authors investigate the effect of inter-NF dependencies in terms of resource consumption in a Free5GC network deployed in Kubernetes. However, only limited monitoring (e.g., RAN parameters are not considered) is undertaken through custom Python scripts. Finally, \cite{li20215growth} introduces the 5GROWTH service platform with an AI-driven automated 5G end-to-end slicing. However, no 5G core is deployed.

% Why this paper is different
Unlike the previous works, the framework proposed in this paper jointly provides full cloud-native deployment of both 5G core and end-to-end monitoring (including RAN) through CNFs orchestrated via Kubernetes. Besides, the framework is visually demonstrated through two scenarios reflecting a series of events in the context of UPF re-selection and user mobility.

%%%% SoA - Papers not included

%Authors in \cite{wiranata2020automation} proposes network automation using Kubernetes to manage 5G services. TERRIBLE PAPER. CONS: Mosaic and OAI in Kubernetes but whole core network in one pod. No monitoring

%Authors in \cite{chima2020context} propose a custom Kubernetes scheduler for edge-native 5G applications 5G. CONS: 5G NSA (unknown project) with no cloud-native deployment of 5GC NFs. No monitoring.

%\cite{bolivar2018deployment} depicts an architectural blueprint of 5G-aware open-source evaluation testbed. CONS: no 5G core discussed nor deployed, and not monitoring.

%\cite{perez2021performance} presents a performance comparison of well-known virtualization technologies applied to deploy a real monitoring architecture for multi-site 5G platforms. CONS: No core deployed + RAN monitoring is not considered.

%%%%%%%%%%%%%%%%%%%%%%%%%%%%
%% PLATFORM
%%%%%%%%%%%%%%%%%%%%%%%%%%%%
\section{Cloud-native 5G experimental platform} \label{section:platform}

% 5G Architecture and intro to section
The 5G architecture consists of two parts that have remarkably changed from previous generations: the new radio access network (NG-RAN) supporting the new radio (NR) and the 5G Core Network (5GC). In this section, we describe the main components of our experimental platform and propose a MEC-enabled testbed to demonstrate the end-to-end monitoring system.

%As for the core, 3GPP defines the Service-Based Architecture (SBA)~\cite{3gpp_sba}, which enables modularization of network functions and aligns well with the Network Function Virtualization (NFV) and Software-Defined Network (SDN) principles. Together, these provide agility and flexibility in terms of resource placement and efficient resource utilization, reducing time-to-market for new services and thus helping operators stay or get competitive.

%%% Figure testbed
\begin{figure}[t]
\centering
\includegraphics[width=.86\linewidth]{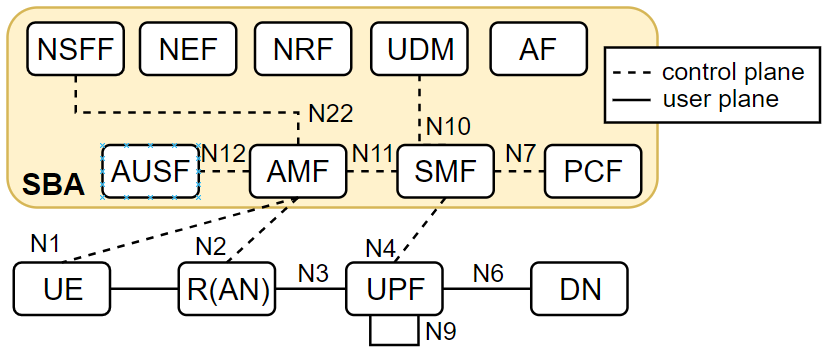}
\caption{5G System Architecture in reference point representation (based on \cite{3gpp_sba}). Some NFs and interfaces are not included for the sake of representation.}
\label{fig:sba}
\end{figure}

%%%%%%%%%%%%%%%
%% 5G CORE
%%%%%%%%%%%%%%%
\subsection{Open-source 5G core}

% 5G core SBA
The movement toward softwarization of telecommunication networks has deeply influenced the creation of the 5G core (5GC)~\cite{5gamericas}. Rather than relying on monolithic elements, 5G adopts a service-based architecture (SBA) composed of NFs that modularize the tasks of the core. As shown in Fig.~\ref{fig:sba}, these NFs interact through Service-Based Interfaces (SBIs), which employ Representational State Transfer (REST) interfaces.
% Comment on slicing
A key feature of such SBA modularization is network slicing, which benefits from softwarization and cloudification. In essence, slices represent logical instances of the network that can be tailored to optimize services and thus cope with different service level agreements (SLAs) according to the use case. Further, within the SBA, we may find the NWDAF (Network Data Analytics Function) and the MDAF (Management Data Analytics Function) for generating insights from NFs data and taking actions to enhance performance, including slice selection and control~\cite{pateromichelakis2019end}.

%\sergio{[JM] si hi ha temps i creus que pot ser d'interès, potser es podria veure quina és la filosofia de monitorització de determinats blocs de la SBA a nivell de monitorització i dir on apliqui que el SW que tu has fet per monitoritzar la RAN s'inspira en la funcionalitat de NWDAF o MDAF o el que encaixi millor. Però el tema és que aquest són molt enfocats a analytics. S'hauria de mirar l'encaix}

% Open-source 5G
This new architecture allows 5G stakeholders much more flexibility and openness, paving the way for an environment where open-source perfectly suits. In this regard, some alternatives of open-source 5GC are available, such as OpenAirInterface~\cite{nikaein2014openairinterface} CN, Free5GC~\cite{free5gc}, and Open5GS~\cite{open5gs}. In our experimental framework, we use Open5GS v2.4.0, since it includes most of the 5GC NFs defined in 3GPP and also allows deploying more than one UPF instance, thus supporting MEC-enabled networks. Nevertheless, our experimental framework is designed to also work with other open-source 5G cores under acceptable adjustments.\footnote{\newver{Preliminary deployments of our containerized framework with OpenAirInterface CN and Free5GC have been already successfully validated.}}

%%%%%%%%%%%%%%%
%% Cloud-native 5G CORE
%%%%%%%%%%%%%%%
\subsection{Cloud-native deployment of the 5G core}

% Why CNFs in industry
5G is expected to support use cases that go beyond raw throughput performance, where the focus is to be put on service flexibility and agility. In this regard, the procedure to deploy 5G NFs has a critical impact. In essence, these functions can be instantiated as physical NFs (PNFs), virtual NFs (VNFs), or containerized NFs (CNFs). Naturally, VNFs gained momentum against siloed PNFs since the conceptualization of the modularized 5G SBA because of the virtualization benefits in terms of efficiency, scalability, or cost. Recently, it is the turn of Containerized Network Function (CNFs) to gain momentum among operators~\cite{chun2019kubernetes} against conventional Virtualized Network Function (VNFs) due to their higher degree of scalability, efficiency for operation and management, energy-saving, and suitability for resource-constrained edge applications.

% What are containers
According to Docker~\cite{docker}, a container is a standard unit of software that packages up code and all its dependencies, so the application runs quickly and reliably from one computing environment to another. This makes a container image a lightweight, standalone, executable package of software that includes everything needed to run an application.
% Orchestrators - why Kubernetes
Container deployments, which may span multiple hosts, are managed with an orchestrator responsible for automating container creation, deletion, and modification without service disruption. Notice that such tasks on containers match the NFV lifecycle management. In this work, we adopt Kubernetes~\cite{kubernetes} as a container orchestrator, since it is the de facto solution in multiple industries for high-demand services with complex configurations.
% Only the sum of requested resources in each node of a cluster is taken into consideration by the default scheduler in Kubernetes. This is not effective enough when resource optimization should also account for potential sudden and drastic performance degradation~\cite{chima2020context}.

% What is a Kubernetes cluster
%A Kubernetes cluster consists of at least one master and multiple computer nodes. The master is responsible for ex-posing the application program interface (API), scheduling the deployments and managing the overall cluster. The smallest unit in Kubernetes is a Pod, which consists of one or more containers that share the same context and resources. Also, each node runs a container runtime, such as Docker, along with kubelet, a Kubernetes agent for maintaining the local pods according to the information provided by Kubernetes API.

\begin{figure}[t]
\centering
\includegraphics[width=.8\linewidth]{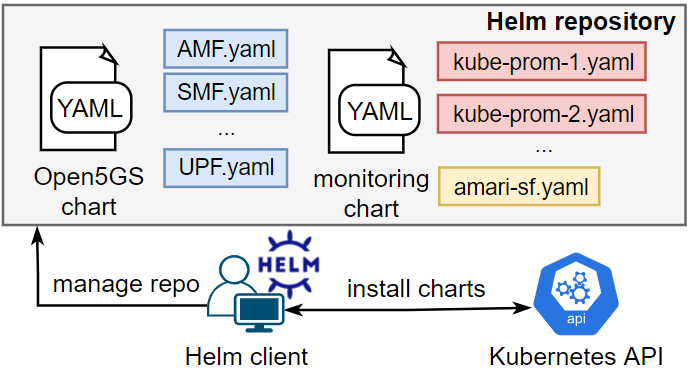}
\caption{Helm and Kubernetes flowchart. The Open5GS chart is composed of the NF templates in blue (e.g., AMF), whereas the monitoring chart includes templates for kube-prometheus in red and the Amarisoft sampling function in yellow.}
\label{fig:helm}
\end{figure}

% Deploying Open5GS and monitoring in Kubernetes
As explained in the following subsections, we deploy both the 5GC NFs and monitoring system within the same Kubernetes cluster using Helm charts~\cite{helm}, a collection of files that describe a related set of Kubernetes resources (see Fig.~\ref{fig:helm}).\footnote{Notice that all the NFs in Open5GS can be compiled and deployed separately, making it a suitable candidate for evaluating the performance of distributed and cloud-native deployments of the 5GC.} This way, the whole framework is deployed in just two commands, one for installing the monitoring system and the other for installing Open5GS. 
% CNI
Once the deployments are instantiated, connectivity among containers and toward external services must be provided. In particular, our Kubernetes cluster relies on Calico~\cite{calico}, a well-known container network interfaces (CNIs) plugin to implement such networking capabilities.

% How to containerize Open5GS
%\sergio{[If there is space left, we may explain how to containerize Open5GS but it just seems so technical/tutorial stuff to me for this paper.]}

%%% Figure K8s

%%%%%%%%%%%%%%%
%% RAN
%%%%%%%%%%%%%%%
\subsection{Experimental RAN integration}

% Intro to RAN
The radio access network (RAN) is another critical component of 5G networks, since it provides individual users with wireless connectivity to the core and external data networks.
% Types of RAN
There are different alternatives when it comes to experimenting with 5G RAN, which can be classified in simulated/emulated (e.g, UERANSIM~\cite{ueransim}) and physical/real (e.g., Amarisoft). In this work, in order to provide actual over-the-air-transmissions, we rely on Amarisoft's AMARI Callbox Ultimate~\cite{amari} acting as a gNB with high-performing NR capabilities. Nevertheless, other RAN alternatives like UERANSIM have been also successfully integrated within the testbed without requiring any configuration changes in the Open5GS Helm chart.

Notice that the RAN in our framework is essentially a physical NF (PNF), whereas the 5GC and monitoring system are built on CNFs. As for how to integrate the AMARI gNB with the Open5GS core, we shall indicate the AMF's IP address in the gNB configuration file to establish the NG Application Protocol (NGAP) connection. In our case, since the AMF runs as a CNF, a custom Kubernetes service exposes the AMF functionality to let the gNB point to the master node IP rather than to the AMF's pod IP. This is a common practice, since pod IPs may change after deletion, while services remain fixed.

%%%%%%%%%%%%%%%
%% Scenario
%%%%%%%%%%%%%%%
\subsection{A MEC-enabled 5G testbed}   \label{sec:testbed}

\begin{figure}[t!]
\centering
\includegraphics[width=\linewidth]{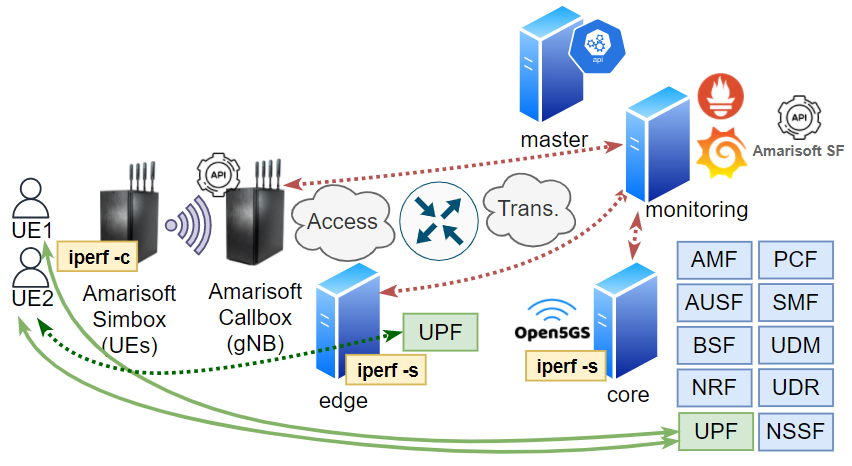}
\caption{MEC-enabled 5G testbed. Data planes are shown in green (dashed arrow for the MEC data plane). Monitoring connections are represented with red dashed arrows. }
\label{fig:marsal-csndsp-testbed}
\end{figure}

% Intro to testbed
Testbeds are essential in telco research, as new architectures, techniques, and features can be conveniently assessed and validated in the lab before going into field trial campaigns.
% Describe testbed - components and nodes
In this work, we implement and integrate the testbed shown in Fig~\ref{fig:marsal-csndsp-testbed} to showcase our cloud-native end-to-end 5G experimental platform. In particular, the 5G network is composed (from left to right) of the following elements: two UEs emulated with Amarisoft AMARI UE Simbox Series~\cite{amari_simbox}, with UE1 always targeting best-effort services while UE2 may target both best-effort and time-critical MEC applications (e.g., AR/VR); an Amarisoft Callbox acting as a stand-alone 5G gNB; an edge node running the MEC UPF and an iperf~\cite{iperf} server; a core node running the Open5GS CNFs and another iperf server; a monitoring node hosting the monitoring containers; and a master node managing the Kubernetes cluster. %\sergio{[Comment that the access and transport networks are just for representation. Actually, there is no difference between them. To think how to explain it.]}
In the current testbed implementation, since the focus is on the monitoring of the RAN and core domain components, the access and transport networks have been simplified to Ethernet links in a local area network. As shown in the Kubernetes representation in Fig.~\ref{fig:k8s}, the cluster is composed of the aforementioned nodes (master, core, edge, and monitoring). Therefore, the master is responsible for deploying, controlling, deleting, and updating the containers of each of the nodes.

%%% Figure testbed

% Open5GS NFs - Multi UPF
Remarkably, all the 5GC NFs are deployed in one click (through Helm) in the core and edge host. In particular, the core runs all the Open5GS NFs available in v2.4.0, including both critical (e.g., AMF or SMF) and secondary NFs (BSF). Besides, for the sake of enabling MEC platforms, the presented cloud-native deployment provides two UPFs: one located at the core and the other, at the edge node. This way, each UPF serves as a PDU session anchor and provides a connection point to different access networks, one being the conventional Internet, and the other any data network that can benefit from MEC processing capabilities, like AR/VR application components.

% Monitoring
Finally, the monitoring node is in charge of running all the containers related to the end-to-end monitoring of the components. To that aim, and as explained in \S\ref{sec:monitoring}, kube-prometheus~\cite{kube-prom} is deployed for monitoring Kubernetes elements, while a custom sampling function is developed to pull metrics from the Amarisoft Callbox API. So, these monitoring CNFs can be viewed as 5G's application functions (AFs) within 5G's SBA.

%%%%%%%%%%%%%%%%%%%%%%%%%%%%
%% E2E Monitoring
%%%%%%%%%%%%%%%%%%%%%%%%%%%%
\section{End-to-end monitoring: from core to RAN} \label{sec:monitoring}

% Intro to Section. Emphasize end-to-end monitoring, cloud-native
We depict below the monitoring system designed for our 5G experimental framework. We shall emphasize two main characteristics that make it a valuable asset: it is cloud-native, meaning that specific CNFs are deployed for monitoring purposes, and it is end-to-end, meaning that deployed CNFs pull metrics both from the core and the RAN domains. The data is gathered into a centralized database where metrics are then plotted in dashboards. 
% Why new monitoring needed

%%%%%%%%%%%%%%%
%% Monitoring components
%%%%%%%%%%%%%%%
\subsection{Monitoring with Prometheus}

%%% Figure K8s
\begin{figure}[t!]
\centering
\includegraphics[width=\linewidth]{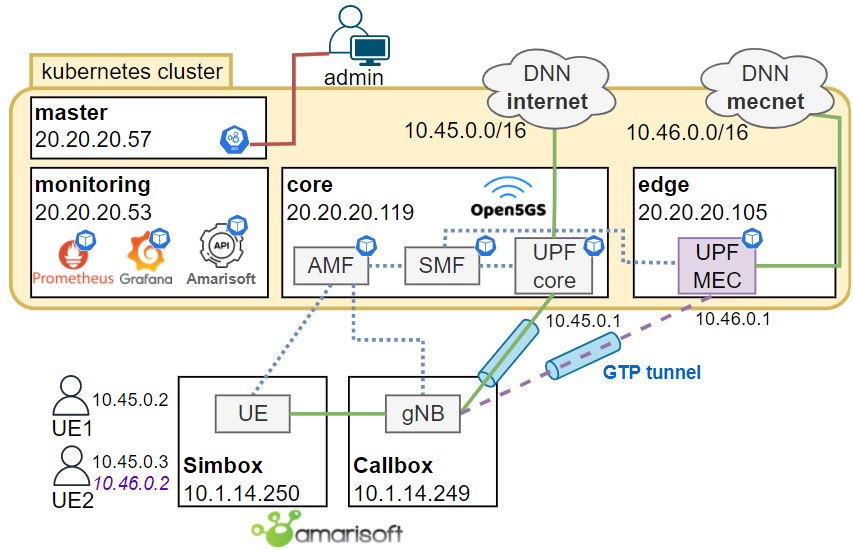}
\caption{Kubernetes deployment. Notice that UE2 gets an IP address depending on its assigned UPF, i.e., 10.45.0.3 and 10.46.0.2 for the core UPF and MEC UPF, respectively.}
\label{fig:k8s}
\end{figure}

%%% Figure grafana
\begin{figure*}[h!]
\centering
\includegraphics[width=\linewidth]{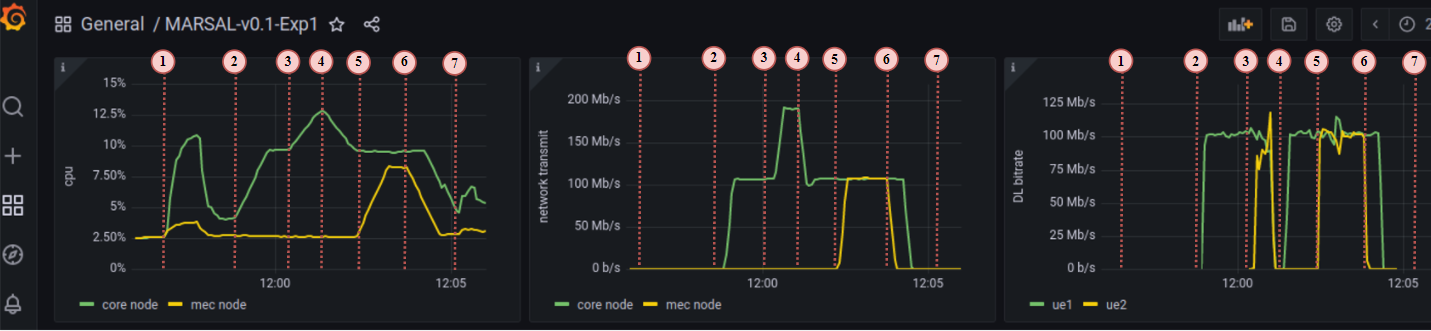}
\caption{Grafana dashboard for experiment \#1 (UPF re-selection). Events are numerated within red circles.}
\label{fig:exp_1}
\end{figure*}

% Use of monitoring
Monitoring is the practice of examining behavioral data from infrastructure, network events, and user interactions. Thus, in order to let network administrators gather metrics of interest and manage unexpected events, a proper monitoring system must be able to collect data from multiple sources.

% Explain why Prometheus
In this work, we primarily rely on Prometheus~\cite{208870}, since it is an open-source monitoring and alerting toolkit that can be easily integrated with Kubernetes to support automatic deployment, letting agents be automatically discovered via service discovery. The Prometheus server also opens interfaces to third-party applications, like web UI or Grafana. We refer the reader to~\cite{tseng2019re} for a comparison on monitoring tools focused on 5G networks.
% How we use Prometheus
In particular, we use Prometheus in two different approaches to tackle infrastructure and RAN monitoring. For the former, we use kube-prometheus~\cite{kube-prom} and rely on a custom sampling function for the latter.

%%%% Infra
\subsubsection{Kube-prometheus for infrastructure}

% What is kube-prometheus
To monitor the usage of infrastructure resources, such as compute, storage, and network, we rely on kube-prometeus stack, an easy to operate end-to-end Kubernetes cluster monitoring stack that uses Prometheus Operator. The stack is pre-configured to collect metrics from all Kubernetes components -- meaning resources are measured at different levels, such as pod, workspace, Kubernetes node, or host -- and it also delivers a default set of dashboards and alerting rules. Therefore, we can deploy a full off-the-shelf cluster monitoring tool with a single Helm command. Among the metrics gathered with kube-prometheus, we find CPU, memory, transmit/receive networking, etc. Some of them are presented at node level in \S\ref{sec:evaluation-exp1}.

% We modified the scraping interval of Prometheus to pull metrics more frequently (i.e., every 5 seconds rather than 15)

%% RAN sampling function
\subsubsection{Custom sampling function for RAN}

For our monitoring purposes, the Amarisoft Callbox can be accessed through a remote API using the WebSocket protocol, which establishes a persistent connection between the client (Amarisoft sampling CNF in monitoring node) and the server (Amarisoft Callbox itself). This API exposes different metrics at gNB/radio level, including (per user and cell id) uplink and downlink bitrate, modulation coding scheme (MCS), channel quality indicator (CQI), or signal-to-noise-ratio (SNR).
% Custom script as a sampling function
The custom sampling function we developed, available at the referred repository~\cite{barrachina2022zenodo}, is a containerized Python script that opens a WebSocket against the Callbox API and exposes some metrics of interest to the Prometheus scraper. We showcase some of these RAN metrics in \S\ref{sec:evaluation-exp2}.

%%%% Grafana
\subsection{Visualizing metrics with Grafana}

% Explain why Grafana
Once the data is being gathered, it is usually convenient to visualize it to quickly grasp the behavior of the network at different domains. For such a task, we use Grafana~\cite{grafana}, a multi-platform open-source analytics and interactive visualization web application that is also included in the kube-prometheus stack. We modified the corresponding chart with the inclusion of two Grafana dashboard descriptors in JSON format for the experiments in \S\ref{sec:evaluation-exp1} and \S\ref{sec:evaluation-exp2}, respectively.

%%%%%%%%%%%%%%%%%%%%%%%%%%%%
%% EVALUATION
%%%%%%%%%%%%%%%%%%%%%%%%%%%%
\section{Use case evaluation} \label{section:evaluation}

% Intro to section: showcase monitoring system
This section showcases and validates the presented end-to-end monitoring framework by displaying different metrics, including both infrastructure and RAN parameters. It does so through two use case experiments based on the testbed displayed in Fig.~\ref{fig:marsal-csndsp-testbed} and Fig.~\ref{fig:k8s}. The first experiment deals with UPF re-selection in MEC-enabled 5G networks, and the second focuses on RAN (gNB) measurements under UE mobility. \newver{The considered events are listed in Table~\ref{table:events}.}

%%%%%%%%%%%%%%%
%% MEC experiment
%%%%%%%%%%%%%%%
\subsection{UPF re-selection in MEC platforms} \label{sec:evaluation-exp1}

% Events of experiment 1
As for Experiment \#1,  we trigger the following series of events: first (1), the 5GC CNFs are deployed through the installation of Helm charts, and the session management function (SMF) assigns both UEs to the core UPF. Second (2), UE1 starts a 100 Mbps UDP downlink iperf connection from an iperf server located in the core node. Then (3), UE2 initiates an iperf of the same characteristics until (4), where the iperf stops and the SMF re-assigns UE2 to the MEC UPF. A new iperf connection is then started by UE2 pointing to the edge node iperf server in (5) until (6), the moment at which both UEs stop their corresponding iperf connections. Finally, the whole deployment (5GC CNFs) is terminated, and the Open5GS containers are deleted in the core and edge nodes.

% Explain outcome of experiment 1
Fig.~\ref{fig:exp_1} shows the Grafana dashboard used for Experiment \#1 consisting of three panels: two for infrastructure metrics (node CPU and networking transmit) measured at the core and edge nodes, and one for a RAN metric (downlink bitrate) measured at the Amarisoft Callbox acting as gNB. The temporal events are highlighted with red circles. In (1), we observe a peak of CPU caused by the deployment of the 5GC CNFs in the nodes. In (2), CPU and network transmit increase due to the iperf traffic triggered by UE1. In (3), UE2's traffic raises the CPU and transmit networking of the core node, since UE2 is also assigned to the core UPF. Instead, when UE2 is assigned to the MEC UPF (5), CPU and networking resources are shared between the core and edge node. Finally, we observe a peak on CPU in (7) corresponding to the CNFs termination.

% Conclusion of Experiment 1
This use case shows the potential value of the presented monitoring framework in 5G deployments, where important metrics from different domains (e.g., infrastructure and RAN) can be assessed in an integrated and automated end-to-end manner.

% Please add the following required packages to your document preamble:
% \usepackage{booktabs}
\begin{table}[b]
\caption{\newver{List of events in experiments \#1 (UPF re-selection) and \#2 (UE mobility).}}
\centering
\resizebox{0.46\textwidth}{!}{
\begin{tabular}{@{}ll@{}}
\toprule
Event & Description                                                    \\ \midrule
\#1.1      & 5GC CNFs deployed and UEs assigned to core UPF                    \\
\#1.2      & UE1 starts a 100 Mbps UDP downlink iperf connection            \\
\#1.3      & UE2 starts a 100 Mbps UDP downlink iperf connection            \\
\#1.4      & UE2 iperf stopped and assigned to MEC UPF      \\
\#1.5      & UE2 restarts a 100 Mbps UDP downlink iperf connection          \\
\#1.6      & Both UEs stop their corresponding iperf connections            \\ 
\#1.7      & Whole deployment (5GC CNFs) is terminated                      \\ \midrule
\#2.1      & UE1 starts a 120 Mbps uplink iperf to the core node \\
\#2.2      & gNB reduces the receiver gain 4dB (-4 dB aggregated)           \\
\#2.3      & gNB reduces the receiver gain 4dB (-8 dB aggregated)           \\
\#2.4      & gNB reduces the receiver gain 4dB (-12 dB aggregated)          \\ \bottomrule
\end{tabular}}
\label{table:events}
\end{table}

%%%%%%%%%%%%%%%
%% Mobility
%%%%%%%%%%%%%%%
\subsection{UE mobility}    \label{sec:evaluation-exp2}

\begin{figure}[t]
\centering
\includegraphics[width=\linewidth]{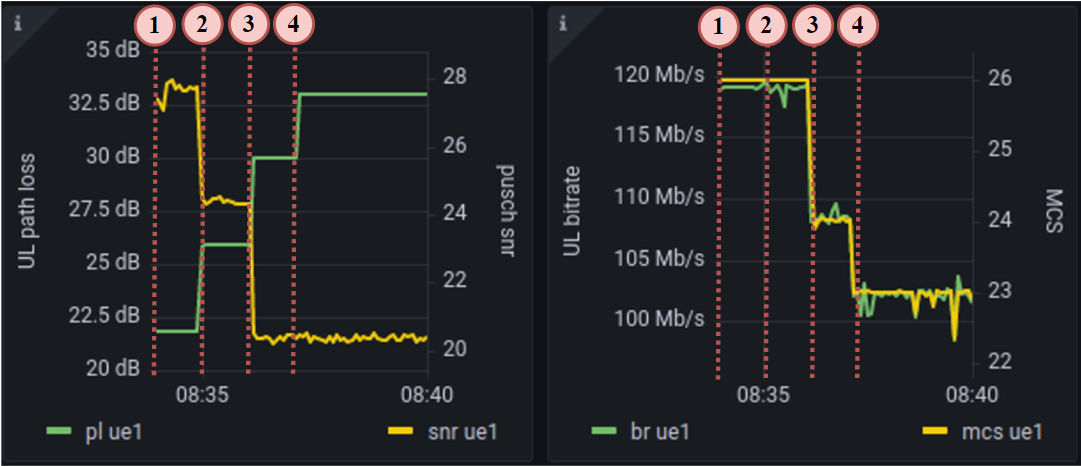}
\caption{Grafana dashboard for Experiment \#2 (UE mobility). Events are numerated within red circles.}
\label{fig:exp_2}
\end{figure}

% Explain experiment 1 including outcome
Finally, to test the containerized Amarisoft sampling function, in Experiment \#2 we focus solely on RAN metrics. To do so, we show in Fig.~\ref{fig:exp_2} a Grafana dashboard corresponding to a scenario where UE1 moves away from the gNB, resulting in higher path loss (lower SNR) and lower MCS and, consequently, lower bit rates. In particular, the series of events is as follows: (1) UE1 starts a 120 Mbps uplink iperf connection to the core node, and from (2) to (4) we sequentially reduce by 4 dB the receiver gain at the gNB through the Amarisoft API to emulate UE1 moving away. As expected, this results in higher path loss and lower SNR at the gNB, consequently achieving lower MCS and lower bitrate.

% Conclusion of Experiment 2
Experiment \#2 has therefore demonstrated the suitability of the developed Amarisoft sampling function in terms of integrating pure RAN metrics into the whole monitoring framework. Providing such end-to-end monitoring is of critical importance to let 5G network administrators control and manage their services, especially when it comes to end-to-end network slicing.

%%% Figure grafana

%%%%%%%%%%%%%%%
%% Alerting
%%%%%%%%%%%%%%%
% \subsection{\sergio{Reaction to alerts}}

% \sergio{JUST AN IDEA: If needed/wanted, we could think of an extra experiment where some alerting is triggered. For instance, if the CPU of the core node exceeds 50\%, some NFs may be moved to another node. RISK: time and unknown outcome.}

% \sergio{I would say this is not necessary since the focus of the paper is not on the algorithmic/intelligence of the network, but on showcasing a cloud-native deployment + end-to-end monitoring.}

%%%%%%%%%%%%%%%%%%%%%%%%%%%%
%% CONCLUSIONS
%%%%%%%%%%%%%%%%%%%%%%%%%%%%
\section{Conclusions} \label{section:conclusions}

In this work, we have implemented, tested, and validated a cloud-native 5G framework with containerized end-to-end monitoring. Using a MEC-enabled 5G testbed with over-the-air transmissions, we depict how to integrate a fully operative 5G framework using CNFs in a multi-node Kubernetes cluster, including an open-source 5G core, a commercial RAN, and an end-to-end monitoring system. We demonstrate the multi-UPF and monitoring capabilities of the framework via Grafana dashboards that display different metrics of interest, gathered both from the infrastructure and radio/RAN domains. Therefore, we expect this paper helps the community in easily deploying 5G monitoring through an integrated and automated end-to-end manner.

% Next steps
As for future work, we are endowing our experimental platform with intelligence so that, e.g., different configurations for the placement of NFs can be adopted as a function of certain events or alerts triggered by the monitoring platform. \newver{Besides, we plan to provide an in-depth comparison of our framework when deployed with other relevant open-source 5G cores.}

% ----------------------------------------------------------------

%% ACKs
\section*{Acknowledgment}

This work has been partially funded by the MARSAL project from the European Union’s Horizon 2020 research and innovation programme under grant agreement No 101017171.

%% Bibliography
\bibliographystyle{IEEEtran}
\bibliography{bib}

\end{document}